# A Rigorous Analysis of Plane-transformed Invisibility Cloaks

Yu Luo[1], Jingjing Zhang[1], Hongsheng Chen[1,2]*, Lixin Ran[1], Bae-Ian Wu[2], and Jin Au Kong[1,2], *Fellow*, IEEE

[1]*The Electromagnetics Academy at Zhejiang University, Zhejiang University, Hangzhou 310027, P. R. China,*

[2]*Research Laboratory of Electronics, Massachusetts Institute of Technology, Cambridge, Massachusetts 02139, USA*

**Abstract**

The electromagnetic characteristics of *plane-transformed* invisibility cloaks are quantitatively studied in this paper. We take elliptical cylindrical cloak as the example, and use an elliptical cylindrical wave expansion method to obtain the scattered field. It is demonstrated that an ideal elliptical cylindrical cloak is inherently visible. Noticeable field scattering and penetration will be induced when the cloak is exposed directly to an electromagnetic wave. However, as long as the cloak consists of a perfect electric conducting lining at the interior surface, perfect invisibility can still be achieved along the direction parallel to the major axis of the cloak for transverse magnetic illumination. Another plane-transformed cloak with a conical geometry is also proposed. The advantage of this cloak is that all the permittivity and permeability elements are spatially invariant while none of them is singular. Hence, it is easily realizable with artificially structured metamaterials. Finally, we show that this kind of cloak can also be used to cloak objects on a flat ground plane.

*Index term*—Invisibility cloak, scattering characteristics, spatial coordinate transformation, artificially structured metamaterials

*Author to whom correspondence should be addressed*; *electronic mail*: chenhs@ewt.mit.edu

# I. Introduction

Design of electromagnetic cloak of invisibility has attracted much attention in the recent years. By exploiting the frequency dispersion of plasmonic materials, Alù and Engheta suggested the possibility of employing plasmonic shell for reducing the scattering from a single or a collection of small spheres [1, 2]. This concept has also been used to conceal a particle simultaneously at different frequency in optical domain [3]. Another methodology based on spatial coordinate transformation of Maxwell's equations was reported by Pendry *et. al* [4]. And a spherical cloak with spatially gradient, anisotropic material parameters was proposed, which can bend and guide the incoming wave smoothly around the cloaked region without disturbing the incident field. Leonhardt described a similar optical conformal mapping method to achieve two-dimensional (2D) invisibility by assuming the short wavelength geometrical limit [5]. The effectiveness of the transformation based cloak design was first validated by full-wave simulation [6] and ray tracing exercise [7], while a practical 2D cylindrical cloak based on the reduced set of material parameters was realized experimentally [8]. Following this approach, some authors considered cloak achieved with a high order transformation [9, 10], aiming to minimize the undesired scattering caused by the impedance mismatched at the cloak's outer boundary. The scattering model of spherical cloak was established in [11, 12]. Further theoretical treatments dealing with the boundary conditions of the point transformed [13] and line transformed [14] cloaks have also been presented. These rapid processes made the transformation a hot topic and triggered more and more study on invisibility cloaks [15-23].

We notice that in most of the above discussions, the inner boundary of the cloak is transformed from a point or a line, which results in extreme material parameters (zero or infinity) at the cloak's interior surface. Recently, the transformation approach was applied to elliptical cylindrical coordinate, and an elliptical cylindrical cloak was achieved correspondingly [24, 25]. It was shown that none of the parameters of the cloak is singular [25], since the inner boundary of the cloak is transformed from a plane in this case. However, the performances of this cloak are not so good, especially for transverse electric (TE) illumination [24]. A similar plane-transformed cloak was presented in [26], which was achieved by combing the segments of cylindrical cloak. And a finite element method was performed to validate the cloaking performance, showing that this cloak can only achieve perfect invisibility for transverse magnetic (TM) plane wave propagating in a definite direction [26]. Since all these facts indicate that the scattering property of the plane transformed cloak is unclear, a rigorous analysis is quite necessary.

In this paper, we study the electromagnetic characteristics of the elliptical cylindrical cloak by rigorously solving Maxwell's equations in elliptical cylindrical coordinate. The general boundary equations are derived. It is demonstrated that an ideal elliptical cylindrical cloak cannot achieve perfect invisibility when it is exposed to an EM wave directly. However, when the inner boundary of the cloak is a perfect electric conducting lining, the odd scattering coefficients are exactly zero for TE illumination, while the even scattering coefficients are always zero for TM illumination. This result indicates that the elliptical cylindrical cloak with a perfect electric conducting lining at the interior surface can still achieve perfect invisibility along the direction parallel to the cloak's major axis when it is illuminated by a TM wave. We have also analyzed how the non-ideal parameters affect the cloaking performance quantitatively. It is found that adding loss to the material parameters can dramatically suppress the backward scattering. All these scattering characteristics can be found in any other plane-transformed cloaks. As an example, we propose another type of plane-transformed cloak which has a cone shaped geometry. Full-wave finite-element method is used to verify the cloaking performance. It is found that apart from achieving one-dimensional invisibility, this kind of cloak has the similar capability with the one proposed by Li and Pendry [27] in that it can give any cloaked obstacle the appearance of a flat conducting sheet. However, the cloak presented here has the advantage that all the material parameters are spatially invariant. Therefore, it is much easier to realize with artificially structured metamaterials, compared with other types of cloaks.

## II. Scattering characteristics of elliptical cylindrical cloaks

To begin with, we first take the elliptical cylindrical cloak as the example, and select the related elliptical cylindrical coordinate ($u$, $v$, $z$), whose relationship with Cartesian coordinate ($x$, $y$, $z$) is given by:

$$x = c \cosh u \cos v, \quad y = c \sinh u \sin v, \quad z = z, \qquad (1)$$

where $c$ is the semi-focal distance. In this coordinate system the constant-$u$ contours represent a family of ellipses, while constant-$v$ contours are hyperbolae, as depicted in Fig .1. To create a cloak, we define a spatial transformation that maps an elliptical cylindrical region $0<u'<U_2$ in the original coordinate ($u'$, $v'$, $z'$) into an annular region $U_1<u<U_2$ in the new coordinate ($u$, $v$, $z$) via

$$u' = f(u), v' = v, z' = z \qquad (2)$$

where $U_1$ and $U_2$ are the inner and outer shell coordinate parameters of the elliptical cylindrical cloak, and $f(u)$ is an arbitrary monotonic function which satisfies the condition $f(U_1) = 0$ and $f(U_2) = U_2$. Under this transformation, a plane ($u' = 0$) is transformed to the cloak's interior surface ($u = U_1$) [24, 25]. And the corresponding relative permittivity and permeability elements of the cloak shell (light brown region in Fig. 1) can be expressed in the transformed physical coordinate system ($u$, $v$, $z$) as follows:

$$\varepsilon_{uu} = \mu_{uu} = \frac{1}{f'(u)}, \quad \varepsilon_{vv} = \mu_{vv} = f'(u), \quad \varepsilon_{zz} = \mu_{zz} = f'(u)\left[\frac{\sinh^2 f(u) + \sin^2 v}{\sinh^2 u + \sin^2 v}\right], \quad (3)$$

and all the off-diagonal terms are zero. For simplicity, we select $f(u)$ as a linear function $f(u) = U_2(u-U_1)/(U_2-U_1)$. Thus, Eq. (3) is reduced to

$$\varepsilon_{uu} = \mu_{uu} = \frac{1}{A}, \quad \varepsilon_{vv} = \mu_{vv} = A, \quad \varepsilon_{zz} = \mu_{zz} = A\left[\frac{\sinh^2 f(u) + \sin^2 v}{\sinh^2 u + \sin^2 v}\right], \quad (4)$$

where $A = U_2/(U_2-U_1)$ is a definite constant. It can be seen from Eq. (4) that both $\varepsilon_{uu}$ (or $\mu_{uu}$) and $\varepsilon_{vv}$ (or $\mu_{vv}$) are spatially uniform, while $\varepsilon_{zz}$ (or $\mu_{zz}$) is dependent on both $u$ and $v$. In Fig. 2, we plot the three components of the material parameters in terms of $u$ for different angular coordinate parameters ($v=0$, $\pi/6$, $\pi/2$). Here the semi-focal distance is selected as $c = 0.2$ m, while the inner and outer shell coordinate parameters are set to be $U_1 = 0.5$ and $U_2 = 1$, respectively. The results show that the permittivity and permeability elements are singular only at two points of the cloak's inner boundary (namely $v = 0$ and $v = \pi$), which is quite different from the point transformed [4] and line transformed [6] cloaks.

The performance of the cloak can be examined by an elliptical cylindrical expansion method. For TE polarized wave, the general wave equation governing the $E_z$ field in the cloak layer can be expressed in the elliptical cylindrical coordinate as

$$\frac{\partial^2 E_z}{\partial u^2} + \frac{\partial^2 E_z}{\partial v^2} + c^2 k_0^2 \left[\cosh^2 f(u) - \cos^2 v\right] E_z = 0, \quad (5)$$

where $k_0 = \omega\sqrt{\mu_0 \varepsilon_0}$ is the wave number of free space. Utilizing the separation of variable method and assuming $E_z(u,v) = R(u)S(v)$, one obtains:

$$\left[\frac{\partial^2}{\partial v^2}+\left(p-c^2k_0^2\cos^2 v\right)\right]S(v)=0, \tag{6}$$

$$\left\{\frac{\partial^2}{\partial f^2}-\left[p-c^2k_0^2\cosh^2 f(u)\right]\right\}R(u)=0, \tag{7}$$

where $p$ is the separation constant. Here, Eq. (6) is ordinary Mathieu equation, while Eq. (7) is modified Mathieu equation. These solutions are known as Mathieu functions [28]. Suppose a TE polarized plane wave with electric field $E^i = \hat{z}e^{ik_0(x\cos\beta + y\sin\beta)}$ is incident upon the cloak at 2 GHz, where $\beta$ represents the incident angle characterized by the included angle between $x$ axis and the incoming wave beam. Thus, the incident fields ($u > U_2$), scattered fields ($u > U_2$), internal fields ($u < U_1$), and the fields inside the cloak layer ($U_1 < u < U_2$) can be described as

$$E_z^i = \sum_{m=0}^{\infty} a_m^{(e)} \text{Je}_m(ck_0, \cosh u)\text{Se}_m(ck_0, \cos v)$$

$$+ \sum_{m=0}^{\infty} a_m^{(o)} \text{Jo}_m(ck_0, \cosh u)\text{So}_m(ck_0, \cos v), \tag{8}$$

$$E_z^s = \sum_{m=0}^{\infty} b_m^{(e)} \text{He}_m^{(1)}(ck_0, \cosh u)\text{Se}_m(ck_0, \cos v)$$

$$+ \sum_{m=0}^{\infty} b_m^{(o)} \text{Ho}_m^{(1)}(ck_0, \cosh u)\text{So}_m(ck_0, \cos v), \tag{9}$$

$$E_z^{\text{int}} = \sum_{m=0}^{\infty} f_m^{(e)} \text{Je}_m(ck_0, \cosh u)\text{Se}_m(ck_0, \cos v)$$

$$+ \sum_{m=0}^{\infty} f_m^{(o)} \text{Jo}_m(ck_0, \cosh u)\text{So}_m(ck_0, \cos v), \tag{10}$$

$$E_z^c = \sum_{m=0}^{\infty}\left\{c_m^{(e)}\text{Je}_m[ck_0, \cosh f(u)] + d_m^{(e)}\text{Ye}_m[ck_0, \cosh f(u)]\right\}\text{Se}_m(ck_0, \cos v)$$

$$+ \sum_{m=0}^{\infty}\left\{c_m^{(o)}\text{Jo}_m[ck_0, \cosh f(u)] + d_m^{(o)}\text{Yo}_m[ck_0, \cosh f(u)]\right\}\text{So}_m(ck_0, \cos v), \tag{11}$$

where $Se_m$ (or $So_m$) is the $m$-th order even (or odd) angular Mathieu function, while $Je_m$ (or $Jo_m$), $Ye_m$ (or $Yo_m$), and $He_m^{(1)}$ (or $Ho_m^{(1)}$) represent the $m$-th order even (or odd) radial Mathieu function of the first, second, and third kinds, respectively. $a_m^{(e)} = i^m \sqrt{8\pi} Se_m(ck_0, \cos\beta) / N_m^{(e)}$ and $a_m^{(o)} = i^m \sqrt{8\pi} So_m(ck_0, \cos\beta) / N_m^{(o)}$. $N_m^{(e)}$ and $N_m^{(o)}$ refer to the normalization factors [28]. $b_m^{(e)}$, $b_m^{(o)}$, $c_m^{(e)}$, $c_m^{(o)}$, $d_m^{(e)}$, $d_m^{(o)}$, $f_m^{(e)}$, and $f_m^{(o)}$ are all undetermined expansion coefficients. Consider the case where the internal region ($u < U_1$) is free space. By applying boundary conditions (continuity of $E_z$ and $H_v$) at the two boundaries $u = U_1$ and $u = U_2$, the undetermined coefficients can be obtained as:

$$b_m^{(e)} = \frac{a_m^{(e)} Je_m'(ck_0, \cosh U_1) Je_m(ck_0, 1)}{Je_m(ck_0, \cosh U_1) He_m^{(1)'}(ck_0, 1) - Je_m'(ck_0, \cosh U_1) He_m^{(1)}(ck_0, 1)}, \qquad (12)$$

$$b_m^{(o)} = \frac{a_m^{(o)} Jo_m(ck_0, \cosh U_1) Jo_m'(ck_0, 1)}{Jo_m'(ck_0, \cosh U_1) Ho_m^{(1)}(ck_0, 1) - Jo_m(ck_0, \cosh U_1) Ho_m^{(1)'}(ck_0, 1)}, \qquad (13)$$

$$f_m^{(e)} = \frac{a_m^{(e)} Je_m(ck_0, 1) He_m^{(1)'}(ck_0, 1)}{Je_m(ck_0, \cosh U_1) He_m^{(1)'}(ck_0, 1) - Je_m'(ck_0, \cosh U_1) He_m^{(1)}(ck_0, 1)}, \qquad (14)$$

$$f_m^{(o)} = \frac{a_m^{(o)} Jo_m'(ck_0, 1) Ho_m^{(1)}(ck_0, 1)}{Jo_m'(ck_0, \cosh U_1) Ho_m^{(1)}(ck_0, 1) - Jo_m(ck_0, \cosh U_1) Ho_m^{(1)'}(ck_0, 1)} \qquad (15)$$

It can be seen from Eqs. (12)–(15) that the scattering coefficients $b_m^{(e)}$ and $b_m^{(o)}$, as well as the penetration coefficients $f_m^{(e)}$ and $f_m^{(o)}$ are non-zero. With the help of Eqs. (8)–(11), the total electric field can be calculated. As shown in Fig. 3, noticeable field scattering and penetration can be observed. Hence, the ideal elliptical cylindrical cannot achieve perfect invisibility in this case.

For the case considered in [23, 24], where a PEC lining is applied to the inner boundary of the cloak, the internal fields are exactly zero ($f_m^{(e)} = f_m^{(o)} = 0$). By applying the boundary conditions, we can determine the scattering coefficients

$$b_m^{(e)} = -\frac{a_m^{(e)} \mathrm{Je}_m(ck_0,1)}{\mathrm{He}_m^{(1)}(ck_0,1)}, \quad b_m^{(o)} = 0 \tag{16}$$

Eq. (16) shows that the odd scattering coefficient $b_m^{(o)}$ is always equal to zero, which indicates that when the incident field is odd symmetrically distributed (namely, $a_m^{(e)} = 0$), the scattered field outside is zero. However, for a plane wave scattering problem, $a_m^{(e)}$ is a nonzero term. Therefore, undesired scattering is still inevitable. Fig. 4 displays the total electric field distribution when the cloak with a PEC lining at the interior surface is subject to TE illumination with different incident angle. Distinct scattering can be observed from the results.

We next consider the problem of TM mode scattering, where the magnetic field *H* is polarized along *z* direction. Suppose a PEC elliptical cylinder is surrounded by an elliptical cylindrical cloak. Following similar processes of TE case, the boundary equations can be listed. By applying the continuity of $E_v$ at $u = U_1$ and $u = U_2$, we get

$$b_m^{(e)} = 0, \quad b_m^{(o)} = -\frac{a_m^{(o)} \mathrm{Jo}_m{'}(ck_0,1)}{\mathrm{Ho}_m^{(1)'}(ck_0,1)} \tag{17}$$

It can be seen that the even scattering coefficient is always zero, indicating that for even symmetrical incident field ($a_m^{(o)} = 0$), zero scattering can be achieved. It is worthwhile to notice that as long as the incident angle $\beta$ is equal to 0 (or π), $a_m^{(o)}$ is 0. In other words, when a TM plane wave is incident upon the cloak along the horizontal direction, perfect invisibility can be realized. Fig. 5(a) shows that the incoming TM wave has been smoothly guided around the internal region without introducing any scattering outside. However, as the incident wave is propagating along any other direction, the scattered field will no longer be zero. As depicted in Fig .5(b), when the cloak is illuminated by a TM plane wave along the vertical direction, large backward scattering will be induced. In fact, this incident angle dependent scattering is caused by the interior surface of the cloak, which is transformed from a plane in the initial coordinate. When the incident direction is parallel to the plane, the tangential electric field at inner boundary of the cloak is zero. Thus, the cloak will have the effect of guiding wave around. However, as the plane wave travels along any other direction, the tangential electric field is no longer zero at the cloak's interior surface. As a result, noticeable scattering will be induced. In order to improve the cloaking performance, one possible solution is to decrease the energy that penetrates into the inner boundary of the cloak. This can be achieved by adding loss to the material parameters [18]. Fig .5 (c) shows the case where a loss tangent 0.05 is introduced to $\mu_z$. It can be seen from the result that the backward scattering has been dramatically reduced.

## III. Plane-transformed cloaks with spatially invariant permittivity and permeability tensors

The full-wave analysis in Sec. II shows that although the elliptical cylindrical cloak can only achieve perfect invisibility for TM illumination in one direction, it has the merit in the easier realization in practice, since the extreme value of the material parameters at cloak's interior surface can be avoided. However, from Eq. (4), we can see that $\varepsilon_{zz}$ and $\mu_{zz}$ are still inhomogeneous. In order to further simplify the material parameters, in this section we propose another plane-transformed cloak, whose permittivity and permeability elements are all spatially uniform. The 2D schematic and 3D geometry of this cloak are depicted in Fig. 6 (a) and (b), respectively. The associated spatial distortion can be described by the following mapping:

$$\rho' = \rho,\ \varphi' = \varphi,\ z' = \frac{H_2}{H_2 - H_1}\left(z - \frac{R-\rho}{R}H_1\right),\ \text{for } z > 0;$$

$$\rho' = \rho,\ \varphi' = \varphi,\ z' = \frac{H_2}{H_2 - H_1}\left(z + \frac{R-\rho}{R}H_1\right),\ \text{for } z < 0. \qquad (18)$$

As shown in Fig. 6, a circular plane is transformed to two closed conical surfaces (which correspond to the inner boundary of the cloak). And the cloak's outer surface is perfectly matched to the background. It is worthwhile to point out that this plane-transformed conical cloak is different with all the 2D cylindrical cloaks in that it has finite size along the axial direction (z direction). Therefore, it can wrap an obstacle completely inside the cloak shell. Using the methodology introduced in Ref. [4], the relative permittivity and permeability tensors of the cloak can be deduced:

$$\bar{\bar{\varepsilon}} = \bar{\bar{\mu}} = \begin{bmatrix} \frac{H_2}{H_2 - H_1} & 0 & -\frac{H_1 H_2}{(H_2 - H_1)R} \\ 0 & \frac{H_2}{H_2 - H_1} & 0 \\ -\frac{H_1 H_2}{(H_2 - H_1)R} & 0 & \frac{H_2 - H_1}{H_2} + \frac{H_2}{H_2 - H_1}\left(\frac{H_1}{R}\right)^2 \end{bmatrix},\ \text{for } z > 0;$$

$$\bar{\bar{\varepsilon}} = \bar{\bar{\mu}} = \begin{bmatrix} \frac{H_2}{H_2 - H_1} & 0 & \frac{H_1 H_2}{(H_2 - H_1)R} \\ 0 & \frac{H_2}{H_2 - H_1} & 0 \\ \frac{H_1 H_2}{(H_2 - H_1)R} & 0 & \frac{H_2 - H_1}{H_2} + \frac{H_2}{H_2 - H_1}\left(\frac{H_1}{R}\right)^2 \end{bmatrix},\ \text{for } z < 0. \qquad (19)$$

The above material parameters are described in cylindrical coordinate system. It is interesting to see that all the elements are spatially invariant. Additionally, none of them are singular. Due to the irregular geometry of the cloak, here we use full-wave finite element method to validate the cloaking performance. Suppose a plane wave with magnetic field parallel to *x-y* plane is incident upon the cloak with different included angle. The frequency is set to be $f = 2$ GHz (corresponding wavelength $\lambda = 15$ cm). The parameters of the cloak are selected $R = 2\lambda$, $H_1 = \sqrt{3}/3 R$, $H_2 = \sqrt{3}R$. The simulated results are plotted in Fig. 7. It is found that the EM wave with wave number parallel to the *x-y* plane (or electric field polarized along *z* direction) can be guided around the internal region, while EM wave from other directions will cause noticeable scattering outside.

From the above analysis, it can be seen that the plane-transformed conical cloak can make the internal PEC cone appear as a conducting sheet. Thus, it is quite useful in the case where an obstacle is located at a flat ground plane. To confirm this point, suppose the upper half of the cloak is placed on the surface of an infinite PEC sheet. A Gaussian beam is launched at 45° with respect to the surface normal from the left. Fig. 8 depicts the total magnetic field distribution, which is similar to the results obtained in Ref. [22], showing that the cloak can successfully mimic a flat PEC plane.

## IV. Conclusions

In summary, we have quantitatively analyzed the scattering characteristics of the elliptical cylindrical cloak by using an elliptical cylindrical wave expansion method. Though the ideal elliptical cylindrical cloak obtained by plane transformation can only achieve perfect invisibility in one direction for TM illumination, it is much easier to be realized with artificially structured metamaterials in practice, since the extreme value (zero of infinity) of the material parameters can be avoided. Another cloak with even simpler material parameters is proposed, which is also realized by applying plane transformation. These plane transformed conical cloak provides great convenience in the fabrication process due to the permittivity and permeability with all the elements spatially uniform.

## Acknowledgement

This work is sponsored by the National Science Foundation of China under grants 60531020, 60701007, and 60801005, in part by the NCET-07-0750, the Zhejiang National Science Foundation under grants R1080320, the ONR under contract N00014-06-1-0001, the Department of the Air Force under Air Force contract F19628-00-C-0002, and the excellent doctoral thesis foundation of Zhejiang University (08009A).

**Figure Captions**

FIG. 1 (color online) Geometry of an elliptical cylindrical cloak. The transformation media that comprises the cloak lies between the two ellipses (light brown region).

FIG. 2 (color online) Relative permittivity and permeability elements as a function of $u$ for three different angular coordinate parameters (a) $v = 0$ (b) $v = \pi/6$; and (c) $v = \pi/2$. Here the semi-focal distance of the elliptical cylindrical cloak is set to be $c = 0.2$ m. The inner and outer shell coordinate parameters are $U_1 = 0.5$ and $U_2 = 1$.

FIG. 3 (color online) Snapshots of the $z$-directed total electric field distributions for a two-dimensional TE mode scattering from an ideal elliptical cylindrical cloak. (a) The cloak is illuminated by a TE plane wave propagating along $+x$ direction; (b) The incoming TE wave is incident upon the cloak along $+y$ direction. Here the internal region ($u < U_1$) is assumed to be free space.

FIG. 4 (color online) $E_z$ field distributions when an ideal elliptical cylindrical cloak with a PEC ling at the inner boundary ($u = U_1$) is exposed to TE illumination with different incident angle: (a) $\beta = 0$; (b) $\beta = \pi/2$.

FIG. 5 (color online) The $z$-directed total magnetic field distribution when a TM plane wave is incident upon an elliptic cylindrical cloak consisting of a PEC lining at the interior surface. Here four cases are considered: (a) $\beta = 0$, and the material parameters of the cloak are lossless; (b) $\beta = \pi/2$, and a lossless cloak is applied; (c) $\beta = \pi/2$, and a loss tangent 0.05 is applied.

FIG. 6 (color online) (a) 2D schematic of a plane-transformed conical cloak in *x-y* plane. The transformation media that composes the cloak lies between the two conical surfaces (blue region). (b) 3D Geometry of the plane-transformed conical cloak.

FIG. 7 (color online) (a) $H_z$ field distribution when a plane transformed conical cloak is illuminated by a TM plane wave along different direction: (a) The plane wave is propagating along *x* direction; (b) The plane wave is incident along *y* direction. The parameters of the cloak are chosen to be $R = 2\lambda$, $H_1 = \sqrt{3}/3 R$, $H_2 = \sqrt{3} R$.

FIG. 8 (color online) (a) Total magnetic field distribution when a Gaussian beam is incident upon a flat conducting sheet at 45° with respect to surface normal. (b) *H*-field distribution when a conical PEC surface surrounded by the cloak is exposed to the illumination.

FIG1

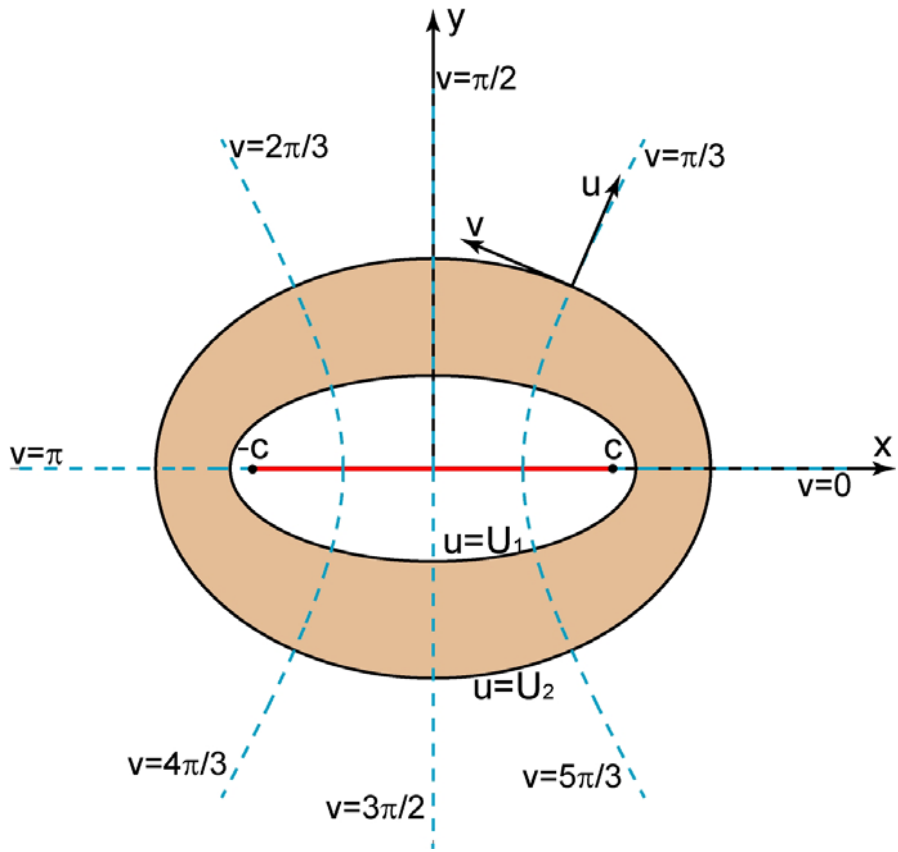

FIG 2

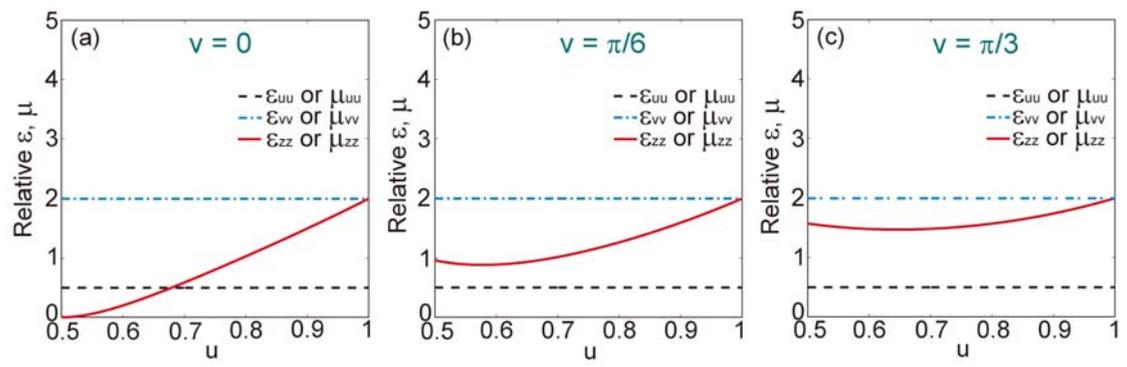

FIG 3

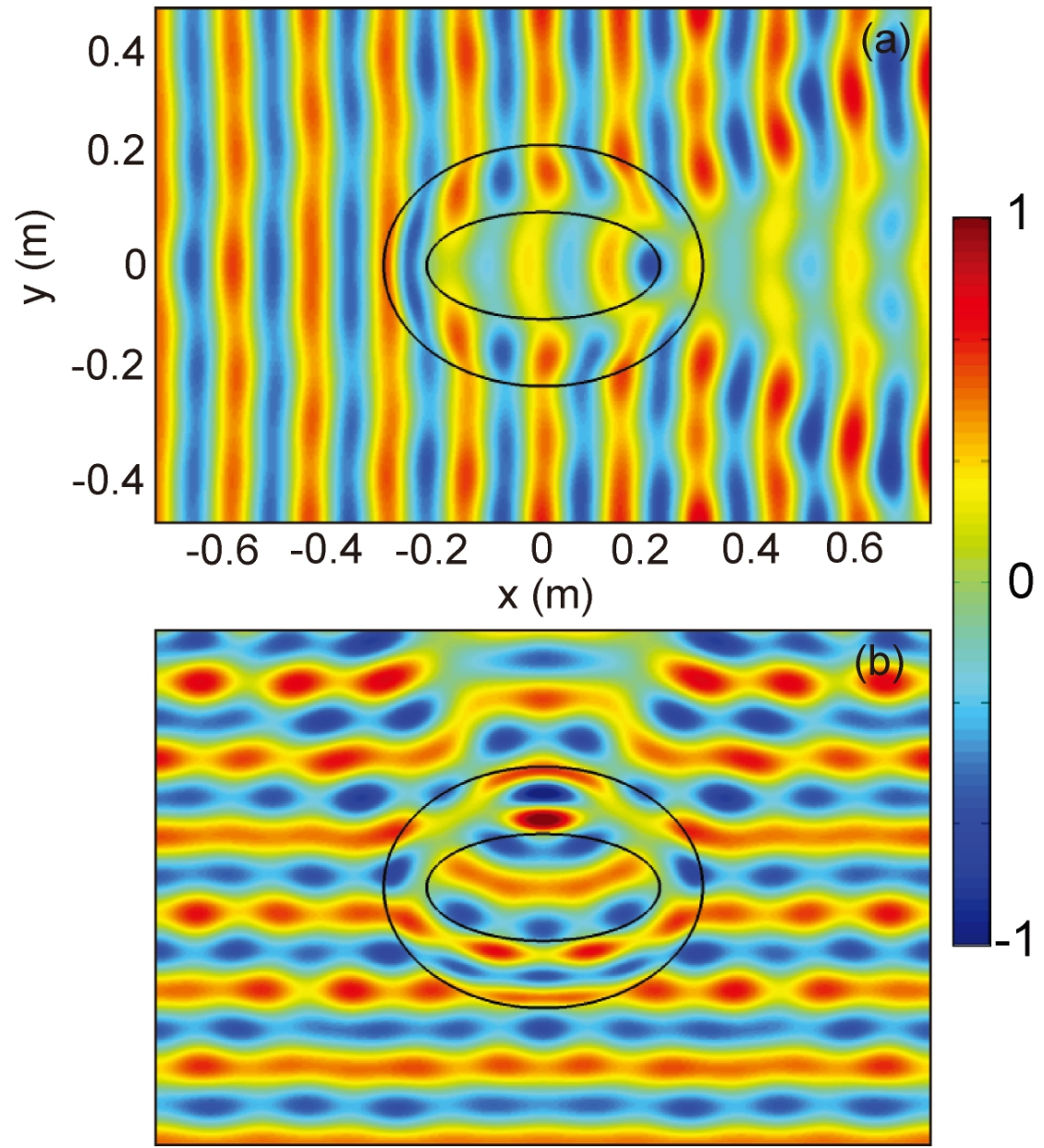

FIG 4

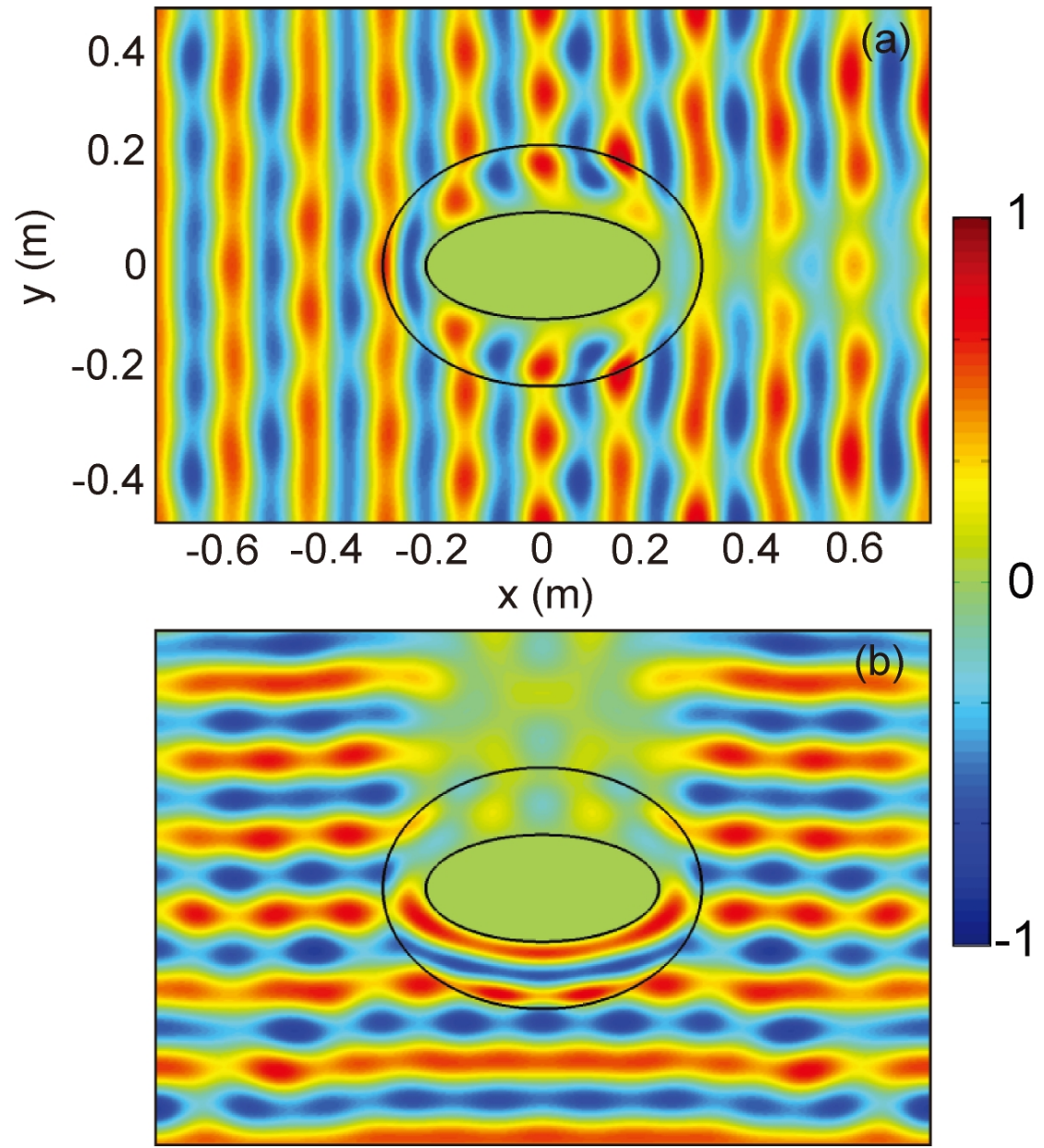

FIG 5

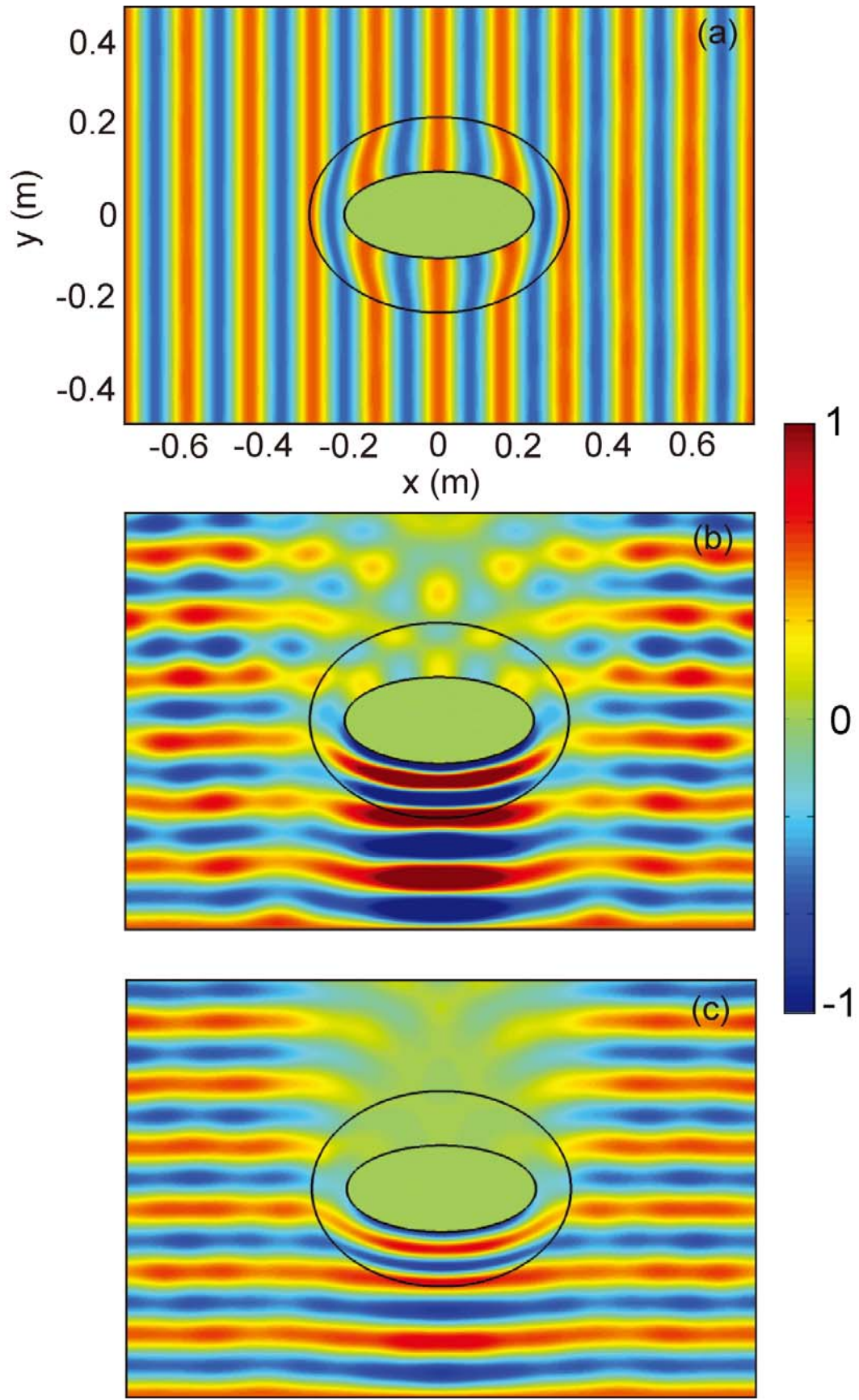

FIG 6

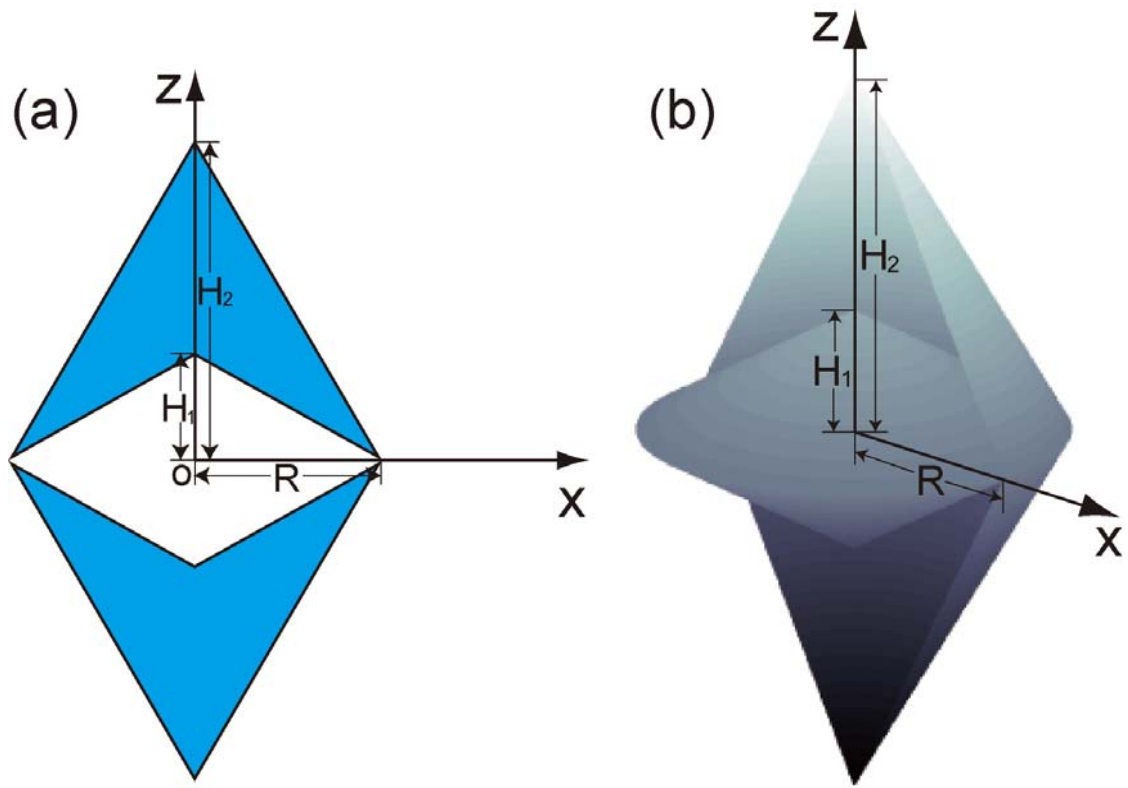

FIG 7

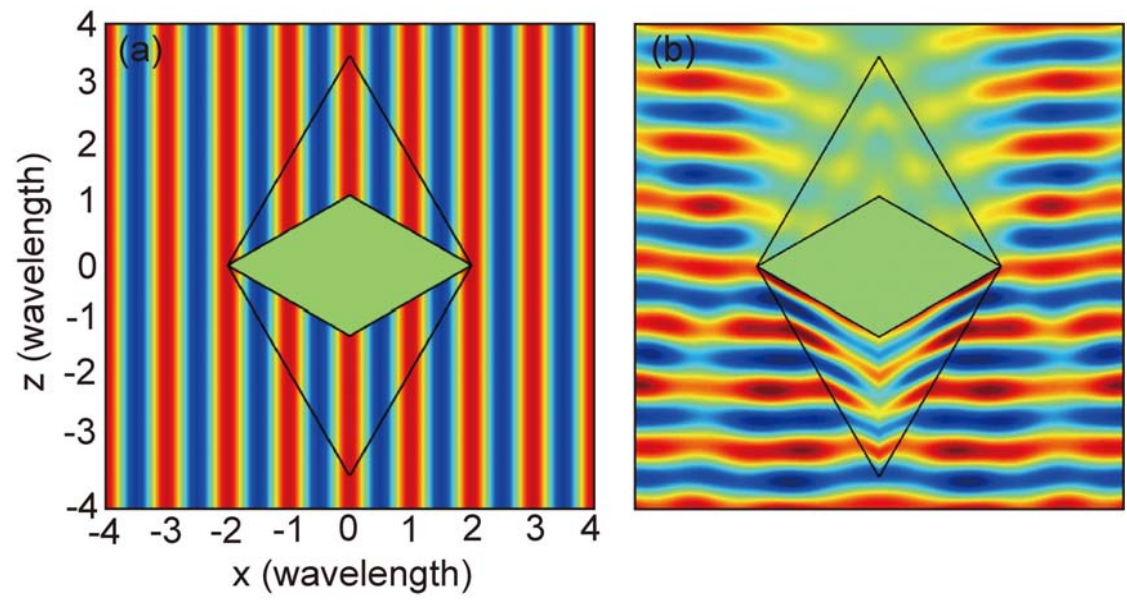

FIG 8

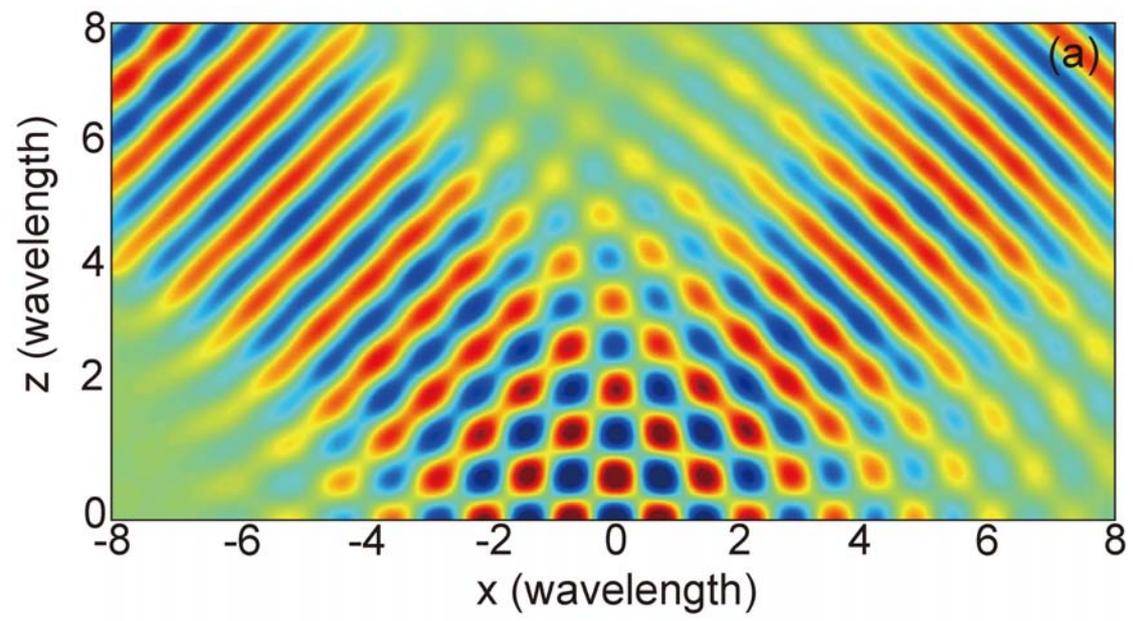

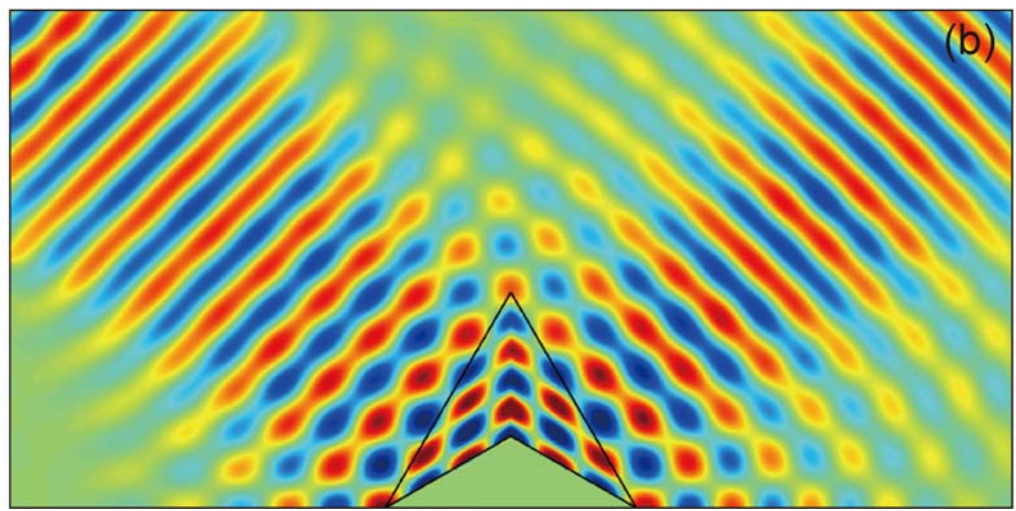